# Temperature dependent phase stability of Mo-Nb-Ta-W refractory high-entropy alloys

Varnita Bajpai[1], Soumyadipta Maiti[1*], Shashank Mishra, Beena Rai

Address: TCS Research, Pune India

* Corresponding author email: soumya.maiti@tcs.com

[1] These authors contributed equally

## Abstract

In this study Mo-Nb-Ta-W refractory high-entropy alloys (R-HEAs) have been studied for their phase stability for a wide temperature range (100 K to 2000 K). The equilibrium thermodynamic phases are determined by the changes in enthalpy and entropy. The enthalpy changes at any temperature are simulated by embedded atom method (EAM) potential based hybrid Monte Carlo molecular dynamics (MC/MD) simulations. Configurational entropy was calculated by quasichemical method. The EAM potentials were all calculated based on the physical input parameters of elements like atomic volume, cohesive energy, elastic constants etc. It was found that the MC/MD evolved equilibrium structures and the degree of local chemical short-range order/clustering (SRO/SRC) largely depend on the ordering enthalpies for various temperatures. The ordering transition temperature is close to 700 K, which is also substantiated by the experimental powder X-ray diffractions done in synchrotron beam. Large increase of degree of next-neighbor B2-type ordering of Mo-Ta pairs and decrease of that Nb-W pairs were observed. This was also expected from the trends of density functional theory (DFT) based *ab-initio* simulations done in the literature and discussed in this work. The simulated diffraction pattern and changes in scattering intensity due to the chemical SRO and SRC was investigated. The diffraction trends were explained with the help of the features of the evolved nanostructure morphology. The developed methodologies may be helpful in prior prediction of long-term phase stability in multi-element alloys by reducing the number of costly experiments.

## 1. Introduction

High-entropy alloys (HEAs) are equimolar/ near equimolar single phase solid-solution alloys of 4 or more elements, which were first discovered by Yeh [1] and Cantor [2] in 2004. Since their discovery the HEAs has been found to be possessing excellent mechanical properties like high strength and ductility at ambient and high temperatures [1, 3, 4], high hardness and wear resistance [4]. The structure of the HEAs are largely found to be a major FCC or BCC solid solution phase stabilized by the enhanced configurational entropy [1-4]. Also, the HEAs have been found to possess various high-performance functional properties like use in diffusion barriers [5], binder for cutting tools [6], cryogenic application [7], enhanced oxidation resistance and hot hardness [8] etc. It has been generally established that observed properties of HEAs are attributed to the locally distorted single-phase lattice structure and sluggish diffusion [3-8].

In 2011 Senkov et al [9] have started using the refractory elements in HEAs which were primarily meant to be used as high temperature sustaining structural materials for aerospace applications. These refractory high entropy alloys (R-HEAs) have been found to be useful in various other applications such as high strength and high toughness materials for both high temperature and ambient temperature [10], material for biomedical use and micro-devices [10, 11], radiation tolerance in nuclear application [12] etc. From an application point of view for these refractory HEAs it becomes important to investigate the temperature and time dependent phase stability for both long-range and short-range order (SRO) changes on the random solid solution phase stability. For example, in the Cr-Ta-W-V R-HEA proposed to be used in futuristic nuclear fusion reactors, SRO/SRC level precipitation happens either by annealing or radiation doses at ambient temperature, which affects their mechanical properties [12]. The TaNbHfZr equimolar R-HEA has been found to undergo SRO with Zr and Hf co-clustering in {1 0 0} habit planes when annealed at 1800°C [11]. At same 1800°C, another MoNbTaW system did not show any experimental evidence of SRO or LRO [10]. This of Mo-Nb-Ta-W based R-HEA has been studied experimentally many times in the literature, but none of the experiments could reveal any occurrence of SRO or LRO in this system. Senkov et al at have done neutron diffraction experiments MoNbTaW R-HEA annealed at 1400°C, but that could not reveal any evidence of local phase-instability [9]. In literature when this R-HEA was synthesized by ball-milling and annealing at 1200°C also did not reveal any sign of local ordering [13]. Thin films of this MoNbTaW alloy deposited by arc deposition and subsequent annealing at various temperature ranges (1000-1400°C) also did not reveal any phase instability [14]. High degree of solid-solution BCC phase stability of this MoNbTaW R-HEA is also observed in additive manufacturing of bulk components as well [15].

The exceptional phase stability of the MoNbTaW however has drawn much interest among the scientific community to theoretically investigate into the local structural ordering of the alloy. Shortly after the R-HEAs were first developed, del Grosso et al had tried to model the next-neighbor based SRO analysis on the MoNbTaW system by the help of Bozzolo-Ferrante-Smith (BFS) method, which involves Equivalent Crystal Theory (ECT) [16]. Subsequently, Widom et al have tested this R-HEA by density functional theory (DFT) based hybrid Monte Carlo Molecular Dynamics (MC/MD) for the SRO structures [17]. The BFS method simulations predicted a higher probability of Ta-Nb next-neighbor (NN) SRO, although the similar electronegativities and atomic sizes of Ta and Nb would otherwise suggest not involving in any B2 type NN SRO [16]. This trend suggested by del Grosso is not supported by the MC/MD simulations of Widom [17]. Widom on the other hand have suggested a high probability of Ta-Mo NN SRO obtained by *ab-initio* DFT based MC/MD, which was not observed by the simulations of del Grosso by BFS method. Later Huhn had done a theoretical study on the order-disorder type A2-B2 (BCC-CsCl) transition in the same R-HEA by an effective Hamiltonian fit with first principles calculations of binary alloy ordering energies [18]. However the predicted A2-B2 ordering temperature found by the method of Huhn was 1654 K, which was not supported by any experiments from literatures where any heat treatment was done close to this predicted transition temperature. Also, this predicted NN ordering temperature from effective Hamiltonian method was very high compared to other estimates from Refs. [16, 17]. Kostiuchenko et al has recently used a DFT based machine-learning (ML) trained potential called low-rank potential (LRP) in combination with Monte Carlo (MC) simulations to calculate the chemical SRO parameters of the MoNbTaW system [19]. Liu at al has done the chemical SRO modeling of the studied R-HEA by DFT based cluster expansion (CE) and ML model development for next-neighbor effective pair interaction parameters (EPI) [20]. In both Refs. [19] and [20], the low temperature SRO parameters show most enhanced NN interaction for Mo-Ta and least interactions for Nb-W pairs, which is in line with the predictions of Refs. [17, 18]. DFT computed EPIs of the MoNbTaW system as studied by Körmann et al [21] shows B2 type strong ordering tendency of Mo-Ta pairs as other references. However, this ordering temperature was varying widely from estimated 1250-700K, as the chemical pairwise interaction range were varying from NN to next to next neighbor (NNN) and so on. Also, in Ref. [19] the degree of SRO parameters and ordering temperature varies significantly depending on whether the corresponding ML models were trained based on relaxed or unrelaxed configurations of DFT. It appears from the existing literature that there can be a lot of variabilities in the predicted degree of SRO parameters and ordering temperature depending on simulation methodologies adopted by various researchers. Moreover, most groups have used hundreds or even thousands of DFT calculations of local configurations with various systems sizes making the existing theoretical approaches of the literature computationally costly and time consuming.

In this study we have investigated the MoNbTaW R-HEA system for its local phase stability both theoretically and experimentally. In the theoretical approach, we have utilized our hybrid Monte Carlo Molecular Dynamics (MC/MD) by employing embedded atom method (EAM) type interaction potentials. In our previous study this kind of simulation technique has shown its potential to predict the local chemical ordering in a multi-component annealed Ta-Nb-Hf-Zr HEA both qualitatively and quantitatively [22]. Temperature dependent chemical SROs of MoNbTaW R-HEA are studied for the anticipated local phase stability at lower annealing temperatures. The theoretical findings are further validated by the X-ray diffraction experiments in synchrotron radiation done on the long-term annealed homogenized samples. The theoretical findings as obtained in this study is compared against the other methodologies and findings in the literature.

## 2. Experiments

At first, elemental powders of Mo, Nb, Ta and W were mixed together and compacted as pellets. These pellets were melted under electric arc in Ar atmosphere. The arc was hold for 2 minutes above the melt. The solidified cast button was flipped upside down 5 times and remolten for a homogeneous melting. The cast button was annealed in 1800°C first for homogenization annealing for 4 days. Again, this material was annealed at long term lower temperature annealing at 500°C for 5 weeks. Any more details of the experimental procedure on the material synthesis can be found in author's previous publication [10].

The long-term annealed samples were pulverized and subsequently put in a 0.5 mm diameter quartz capillary for XRD experiments. Powder XRD pattern of the samples were measured in SLS synchrotron in Paul Scherrer Institute, Switzerland. The energy of the X-ray was chosen as 30keV with a corresponding wavelength of 0.413 Å. The MYTHEN detector was used for data collection with 2θ range of 5-120° and effectively zero background noise. But, in this work the useful diffraction angle up to 40° are plotted and discussed.

## 3. Simulation Methods

An equimolar High Entropy Alloy (HEA) structure of Mo-Nb-Ta-W (25% each) was created in BCC lattice framework. All the atomic sites of the designed 10x10x10 superlattice containing 2000 atoms were distributed randomly to the four elements at first. This random structure was then subjected to annealing under MC/MD at different temperatures of 100 K, 200 K, 300 K, 400 K, 500 K, 600 K,700 K, 800 K, 900 K, 1000 K, 1100 K, 1500 K and 2000K, respectively. NPT simulation were carried out for each temperature at 1atm pressure for 1ns to calculate the lattice parameter at each temperature. A straight line fitted regression model is developed to determine the lattice parameter for any given temperature.

For Molecular Dynamics (MD) simulations, the potential developed for this alloy was of embedded atom method (EAM) type. The EAM is a many-body interatomic potential which consists of the basic formulation in terms of total energy as follows:

$$E_{total} = \sum_i F_i \left( \sum_{j \neq i} f_j(r_{ij}) \right) + \frac{1}{2} \sum_{i,j \, (i \neq j)} \varphi_{ij}(r_{ij}),$$

Where, summation is over all the atoms and $E_{total}$ is the total internal energy of the system, $F_i$ is the embedding function of the atom type at position $i$, $\Phi_{ij}$ is the pairwise interaction function, $f$ is the spherically symmetric electron density function, $i$ and $j$ are the neighboring atoms at the vicinity of each other. The dissimilar type pair interaction is given as follows:

$$\varphi^{ab}(r) = \frac{1}{2} \left( \frac{f^b(r)}{f^a(r)} \varphi^{aa}(r) + \frac{f^a(r)}{f^b(r)} \varphi^{bb}(r) \right),$$

where the superscript $a$ and $b$ denotes atoms of a- and b-types in an alloy system. $\Phi_{aa}$ and $\Phi_{bb}$ are monoatomic potentials given by monoatomic models. $f^a(r)$ and $f^b(r)$ are electron density function for $a$- and $b$-type atoms. In this paper, we have used the Bangwei et al methodology to create the MD potential from initial physical input parameters like lattice constant, cohesive energy, elastic constants and unrelaxed vacancy formation energy [23]. The table of physical input parameters used to make the EAM potentials are given in the following Table 1.

Table 1. Physical input parameters of elements to build EAM potential

| Metal | Lattice Parameter (Å) | Cohesive Energy (eV) | Unrelaxed Vacancy Formation Energy (eV) | C11 (eV/(Å)$^3$) | C12 (eV/(Å)$^3$) | C44 (eV/(Å)$^3$) |
|---|---|---|---|---|---|---|
| Mo | 3.147[a] | 8.691[b] | 3.1[a] | 2.8657[c] | 1.034[c] | 0.699[c] |
| Nb | 3.301[a] | 8.21[b] | 2.75[a] | 1.529[c] | 0.824[c] | 0.177[c] |
| Ta | 3.303[a] | 9.71[b] | 2.95[a] | 1.648[c] | 0.986[c] | 0.516[c] |
| W | 3.165[a] | 11.07[b] | 3.95[a] | 3.263[c] | 1.267[c] | 0.9987[c] |

Table References: a [24], b [25] and c [26].

All the MD simulations were done using LAMMPS package [27]. For Monte Carlo (MC)/ Molecular Dynamics (MD) procedure two random atoms were swapped from their original positions. Then the system was energy-minimized to a convergence level of 10$^{-14}$ eV using conjugate-gradient (CG) method using LAMMPS for the relaxed configurations keeping the LP constant. Then the system was subjected to MC swap where any two random atoms were interchanged from their positions and again energy minimization was done using CG method in LAMMPS package. The change in energy was thus computed before and after the swap. The acceptance criteria followed in this paper was given by Metropolis algorithm as follows:

Accept the move, if $\Delta U < 0$ due to atomic displacements, or
Accept the move with probability $exp(-\Delta U/(k_B T))$, if $\Delta U > 0$,
Where, $\Delta U$ is the change in potential energy, $k_B$ is the Boltzmann constant and $T$ is the temperature.

Again, two atoms were randomly picked, and this process of hybrid-MC/MD was repetitively carried out until the number of attempted swaps reached 50000 swaps. The structures are recorded after every 2000 swaps. This complete procedure is done for all the temperature ranging from 100 K to 1100 K at an interval of 100 K, 1500 K and 2000 K. During the entire simulation the potential energy as well as the coordinates of the system were recorded at intervals.

## 4. Results and Discussions

### 4.1 Thermodynamics

The change in Gibbs free energy of the HEAs due to the MC/MD structure evolutions are related to the change in enthalpy and entropy as

$\Delta G = \Delta H - T. \Delta S$

where, $\Delta G$ is the change in Gibbs free energy, $\Delta H$ is taken from the difference of potential energies of the relaxed configurations of the evolved structure and the initial random structure, $\Delta S$ is the change in entropy between the evolved structure and the initial random structure. In substitutional solid solution alloys, it has been found that that change in vibrational entropy between ordered and disordered phases is under 0.2 $R$/mol/K, which is around one order of magnitude lower than the configurational entropy of formation [28]. Other entropy sources like electronic and magnetic entropies are expected to be tiny for non-magnetic alloy systems [28]. Thus, in this study, focus has been given on the contribution of configurational entropy towards the changes in Gibbs free energy along with changes in enthalpy of the MC evolved systems. In this work the effect of change in configurational entropy from the ideal estimate for random solid solution due to the presence of short-range ordering/clustering (SRO/SRC) is considered. For this, the next-neighbor bond counting statistics for various atomic pair type is applied for the quasichemical model in the following formula

$$S = -R. \sum_{i} x_a \ln(x_a) - \left(\frac{Z}{2}\right) R \left[ \sum_{j} x_{aa} \ln\left(\frac{x_{aa}}{x_a . x_a}\right) + \sum_{k} x_{ab} \ln\left(\frac{x_{ab}}{2. x_a . x_b}\right) \right]$$

Where, $S$ is the entropy, $R$ is the universal gas constant, $x_a$ is the mol fraction of element $a$, $x_{aa}$ is the fraction of next neighbor bonds between the same elements $a$, $x_{ab}$ is the fraction of next neighbor bonds between dissimilar elements $a$ and $b$, $Z=2$ is the coordination factor, respectively [29]. The summation indices $i$, $j$ and $k$ are used to depict the overall composition of individual elements, fraction of next-neighbor bonds of similar atomic species and next-neighbor bond fraction of dissimilar atomic species, respectively.

The change in free energy along with MC/MD structure evolution is plotted for two temperatures (300K and 600K) in Fig. 1. It can be observed from the MC/MD runs for both the different temperatures that, within the initial 5 MC swaps/atom (henceforth called MC swaps) the enthalpy changes rapidly, but the entropic contribution to the free energy is rather insignificant. For 300K and 600K, the entropic contribution becomes rather constant after 15 and 5 MC swaps, respectively. This indicates that the equilibrium free energy is largely determined by the equilibrium enthalpy for various temperature and the local structural changes are enthalpy driven. This was also the trend found for other temperatures of MC/MD simulation as well.

We term the changes in enthalpy of any MC/MD evolved HEA due to local SRO/SRC w.r.t the initial random solid-solution structure as the ordering enthalpy. The variation of this ordering enthalpy w.r.t various temperatures are shown in Fig. 2.

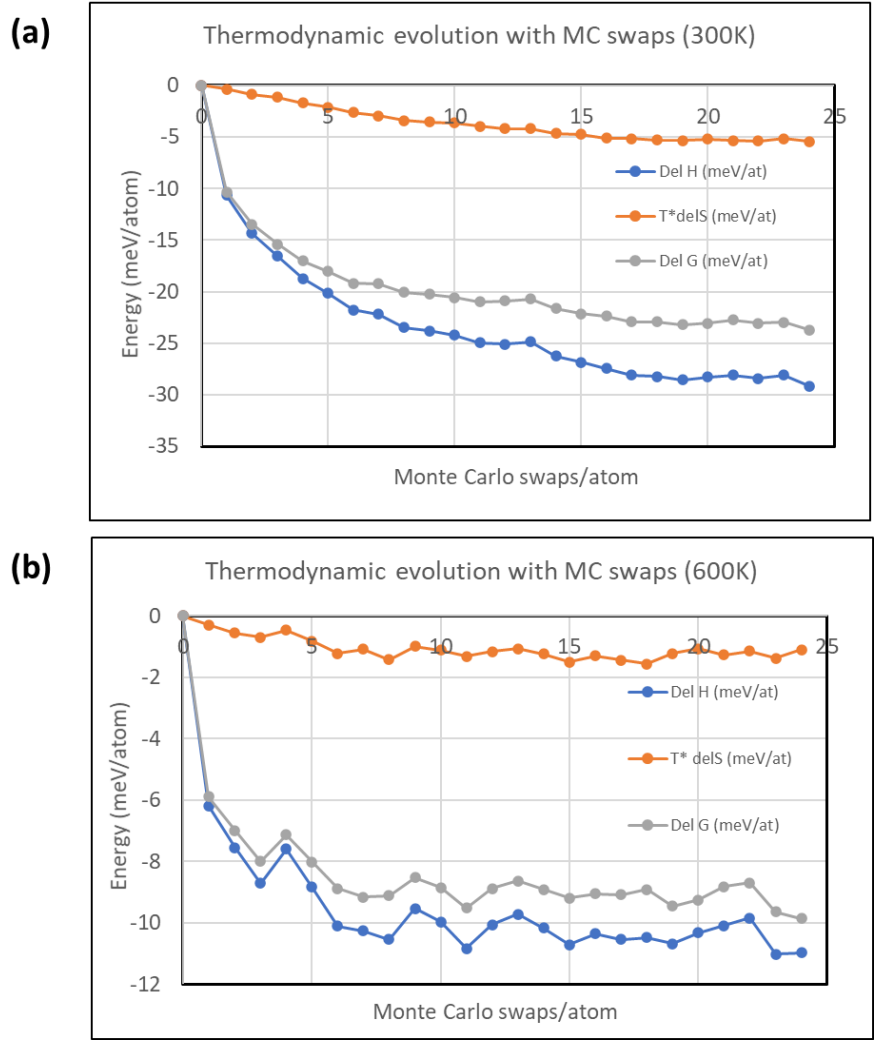

Fig. 1. Changes in thermodynamic properties with isothermal Monte Carlo structure evolution of MoNbTaW HEA; (a) graph for 300K and (b) graphs for 600 K run.

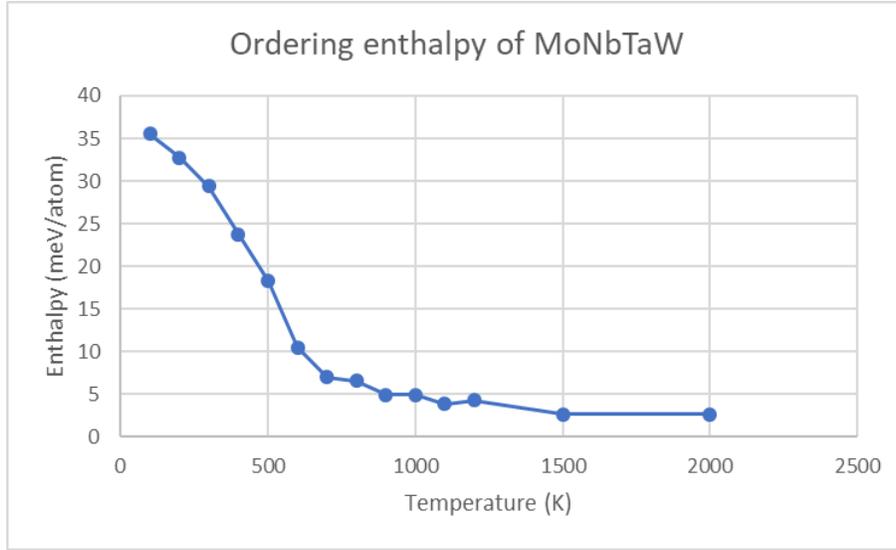

Fig. 2. SRO/SRC local ordering enthalpy w.r.t temperature of MC structure evolution

It can be observed from the Fig. 2 that the significant change in enthalpy due to occurrence of local SRO/SRC becomes significant only below 700 K and it starts to increase below 600 K. From 800 K and above there is hardly any significant change in ordering enthalpy, suggesting that the large thermal fluctuations favoring a random solid solution structure.

**4.2 Chemical Ordering**

The evolution of thermodynamic properties from the above section suggest the occurrence of chemical ordering happening below 700 K. Here we have investigated into the local chemical order by SRO/SRC of the isothermally annealed alloys undergoing Monte Carlo structure evolution at different temperatures. The SRO/SRCs are calculated by using the Warren-Cowley SRO parameters [30]

$$\alpha^{pq} = 1 - \gamma^{pq}/c_p c_q \ ,$$

where, $\alpha^{pq}$ is the SRO parameter for $p$-$q$ element pairs at a particular distance, $\gamma^{pq}$ is the conditional probability of finding an atom of type $p$ at a particular distance given that there is atom type of $q$ at the origin, $c_p$ and $c_q$ are overall concentrations of elements $p$ and $q$ in the system. The calculated order parameters for the 1$^{st}$ next neighbor of the studied HEA are shown in Fig. 3.

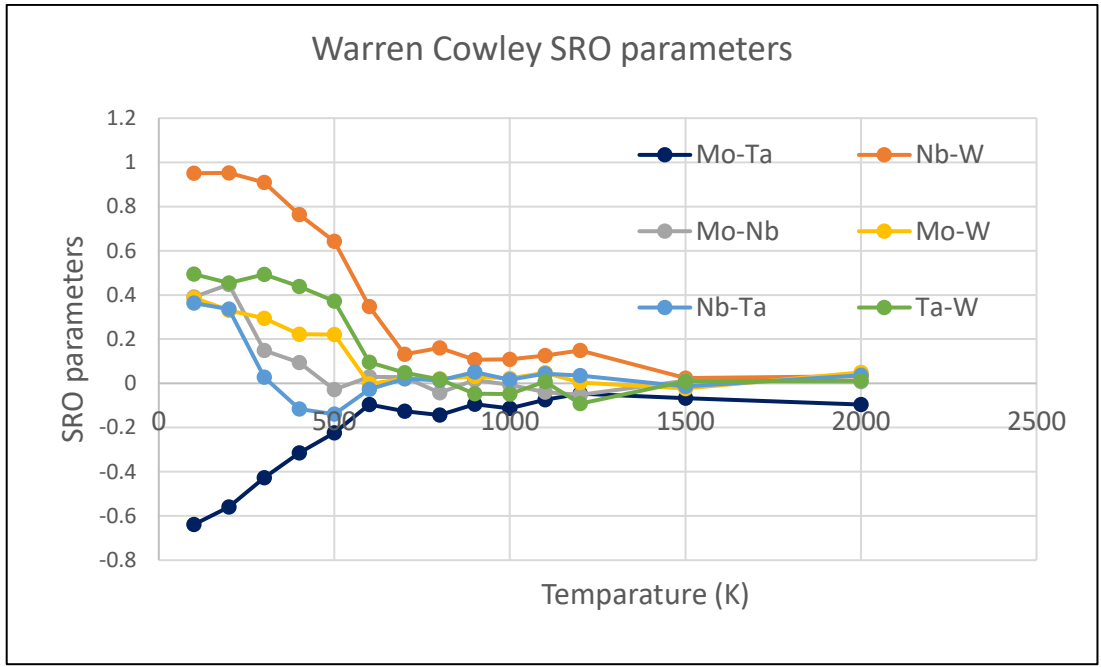

Fig. 3. SRO parameters for next-neighbor elemental pairs of MoNbTaW equilibrium structures for various temperatures.

It can be observed from the trends of the SRO parameter graphs that these parameters start to increase in magnitude from below 600 K. This is also in accordance with Fig. 2 of ordering enthalpy trends, suggesting the changes in enthalpy due to the increased presence of SRO/SRC. It can be also seen that that the SRO parameters remain close to the value of 0 for a wide temperature range from 2000 K to 700 K as expected for a random solid solution alloy. This suggests random solid-solution phase stability for MoNbTaW R-HEA for a very large temperature range.

By looking into the nature of the derived Warren-Cowley SRO parameters, it can be observed that the Mo-Ta next neighbor pair has the strong tendency to order next to each other and Nb-W has the least tendency for next-neighbor ordering. In literature there has been DFT based study on this R-HEA where the ordering enthalpies are calculated for the B2 ordered structure (Cs-Cl type structure) compared to the random solid solution phase [31]. The Mo-Ta ordering has a large negative $\Delta H$ change in enthalpy as -76 meV/atom, largest in magnitude w.r.t all other binary ordering enthalpies of other element pairs. On the other hand, the calculated ordering enthalpy for Nb-W binary is + 9meV/atom. Since the ordered state of Nb-W binary has higher enthalpy levels than their alloyed state, it may be expected that the Nb and W segregate from each other at the next-neighbor level. Whereas large favorable B2 ordering enthalpy of Mo-Ta binary would favor the next-neighbor bonds and opposite for Nb-W bond pairs. This is exactly reflected in the trend found in the SRO parameter analysis in Fig. 3. In the literature Liu et al [20] have developed a DFT based machine learning model for pair interaction based effective Hamiltonian to predict the nature of SRO in the studied MoNbTaW R-HEA system. The effective pair interaction (EPI) parameters as obtained from the ML can effectively capture the SRO effects of the R-HEA. There also the largest EPI parameter is found for Mo-Ta pair and smallest one for Nb-W pairs. It appears that the SRO parameters obtained from the EAM potential based MC/MD calculations are matching with the trends obtained from the DFT based studies of EPI [20] and ordering enthalpies [31]. Other than the nature of the SROs evolving for various temperatures, the determined SRO transition temperature of 700 K for this study is also close to the value of 600 K found by ab-initio DFT based MC/MD technique [17].

### 4.3 Structure Evaluation

In order to investigate into the local lattice structures of the annealed R-HEAs, powder XRD at powerful synchrotron beamline was measured with 0.413 Å wavelength. The powder XRD patterns are shown in Fig. 4a for a high temperature annealed material at 1800°C for 4 days. Another material was subsequently annealed at 500°C for 5 weeks after being annealed at 1800°C for 4 days, and its XRD pattern is given in Fig. 4b.

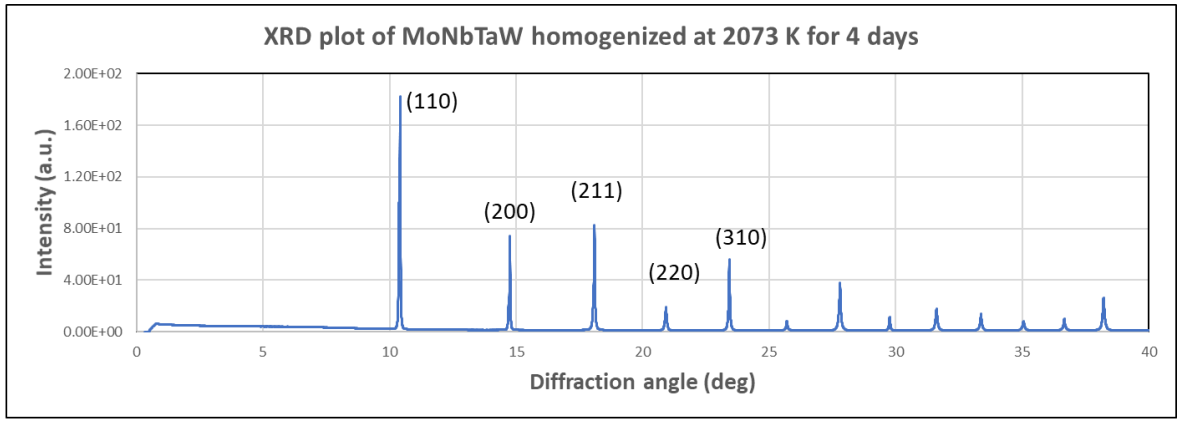

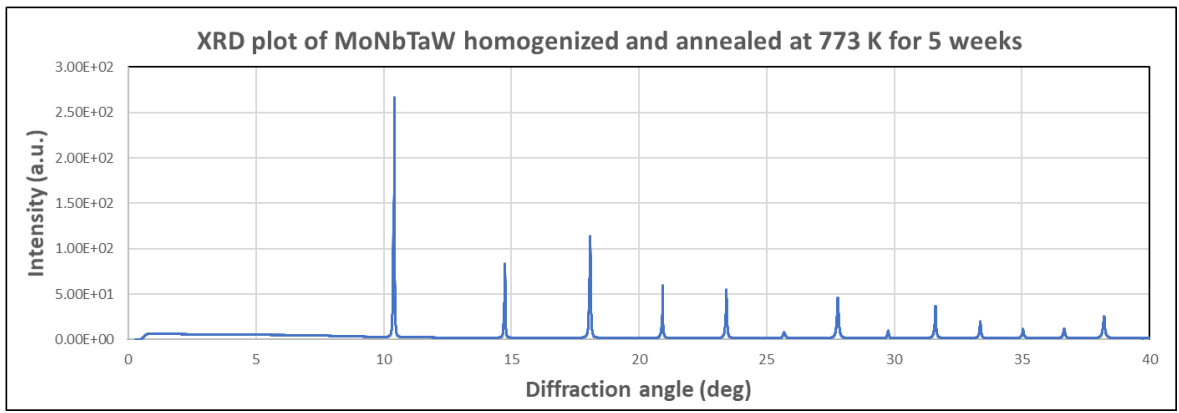

Fig. 4. Powder XRD pattern with monochromatic 0.4Å wavelength beam; (a) pattern for R-HEA annealed at 1800°C, (b) XRD pattern for R-HEA annealed at 1800°C and subsequently annealed at 500°C for 5 weeks.

It can be observed from the above XRD pattern that both the high temperature and low temperature R-HEAs produce a XRD pattern for perfect BCC average structure. For the high temperature annealed structure the observed Bragg reflection peaks are indexed with the BCC XRD peaks as expected from a random solid-solution structure. There is no superlattice Bragg reflections peak observed in the high temperature annealed sample as expected from random solid solution alloys and this is also seen in the other XRD and neutron diffraction studies dome in the literature for samples annealed above 1400°C [9, 10, 13, 14]. To investigate into any occurrence of long-range or short-range local chemical ordering, the 1800°C annealed samples were aged at a much lower temperature of 500°C (773 K) for 5 weeks. The XRD pattern for this long-term aged sample in Fig. 4b however shows no sign of superlattice Bragg reflection peak or diffuse scattering peaks due to local chemical order other than what expected from a BCC random solid-solution phase. These XRD observations from the low temperature aged sample might be expected as the theoretically calculated thermodynamic properties and SRO parameters indicate the presence of any chemical order only below 700 K.

Some MC/MD evolved equilibrium structures are analyzed hereby for their simulated diffraction patterns. This will help us to identify the effects of local chemical order on simulated diffraction and also to compare with the experimental observations. For the simulated X-ray diffraction in the *hk0* reciprocal lattice layer of the MC/MD evolved structures, the following formula was used [32]

$$F(\bm{h}) = \sum_{i=1}^{N} f_i(\bm{h})e^{2\pi i \bm{h} \bm{r}_i} \cdot e^{-\frac{B|\bm{h}|^2}{4}},$$

where, *F(h)* is the structure factor Fourier transform value for the diffraction intensity for the reciprocal lattice vector *h*, $f_i$ is the atomic scattering factor, *B* is the Debye-Waller factor, *N* is the total number of atoms and $r_i$ is the fractional atomic coordinate for the simulated supercell structure. In this study $f_i$ was taken as the atomic number of the atoms for the diffraction calculation. The thermal Debye-Waller factor was kept as zero for the relaxed static structure at 0K with no thermal movement of atoms. This was done to avoid effects from thermal diffuse scattering and highlight the effects from local chemical SRO and structural effects. The magnitude of *F(h)* was plotted in log scale for the grayscale colormap of the simulated *hk0* reciprocal lattice layer as shown in Fig. 5. In Fig, 5a the simulated reciprocal lattice plane of 800 K evolved structure is plotted. It can be observed that even in this low temperature the XRD pattern is like that of random BCC solid solution with no superlattice reflections or diffuse scattering arising from any local chemical order. This is also in accordance with the experimentally observed XRD of the R-HEA isothermally aged at the close temperature of 773 K. However, for the MC/MD evolved structure for 100K there are B2 ordering-like superlattice reflections that can be observed in between the main Bragg reflections. These are shown with red circles in Fig. 5b. These extra reflections are rather diffuse in nature and indicates a SRO type chemical ordering appearing at the nanoscale.

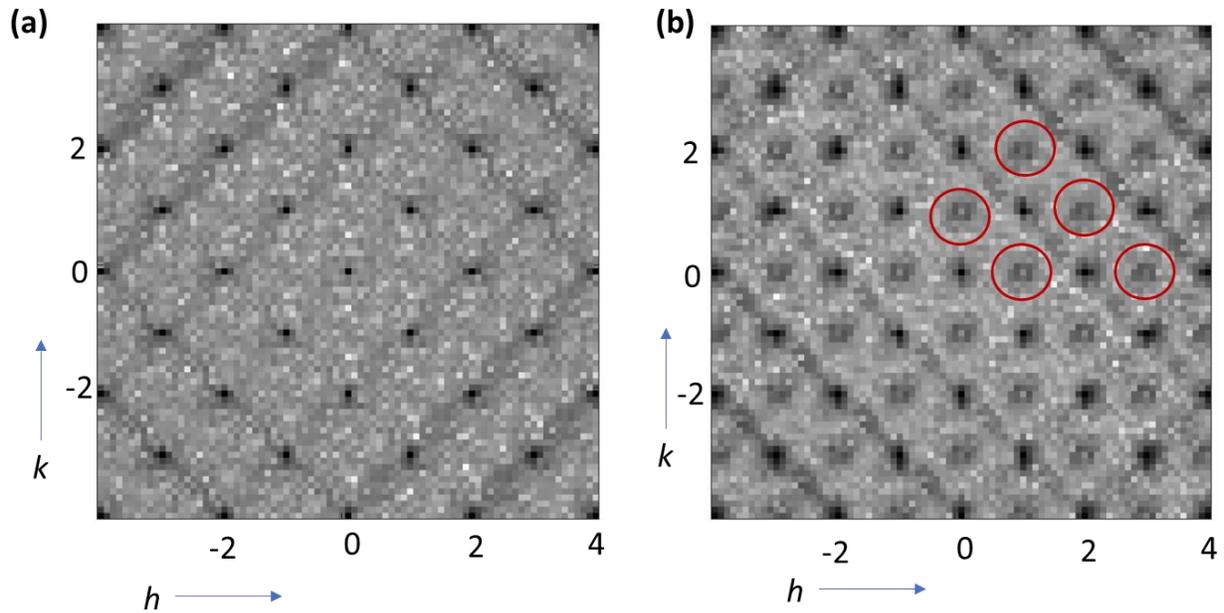

Fig. 5. Simulated *hk0* reciprocal lattice layer for MoNbTaW R-HEA; (a) reciprocal lattice layer for equilibrium structure at 800 K, (b) for structure evolved at 100 K. The extra scattering peaks for 100 K evolved structure is pointed out with red circles.

The powder XRD pattern computed from the reciprocal lattice layers of the two simulated structures are shown in Fig. 6. The X axis is the diffraction vector plotted in the reciprocal lattice unit (r.l.u.). Fig. 6a is for the 800 K MC/MD evolved structure and Fig. 6b is for the 100 K evolved structure, respectively. It can be observed from the simulated diffraction patterns that for the system evolved at 800 K, the pattern look similar to the indexed pattern expected from a BCC solid-solution as well as to the experimentally measured aged sample of 773 K. However, in the simulated pattern of 100 K evolved structure, there is an extra superlattice-type peak appearing as expected from a B2-type local order. This extra ordering peak is shown with an arrow in Fig. 6b. This is appearing because of the extra scattering peaks observed in the reciprocal lattice layer as shown in Fig 5b.

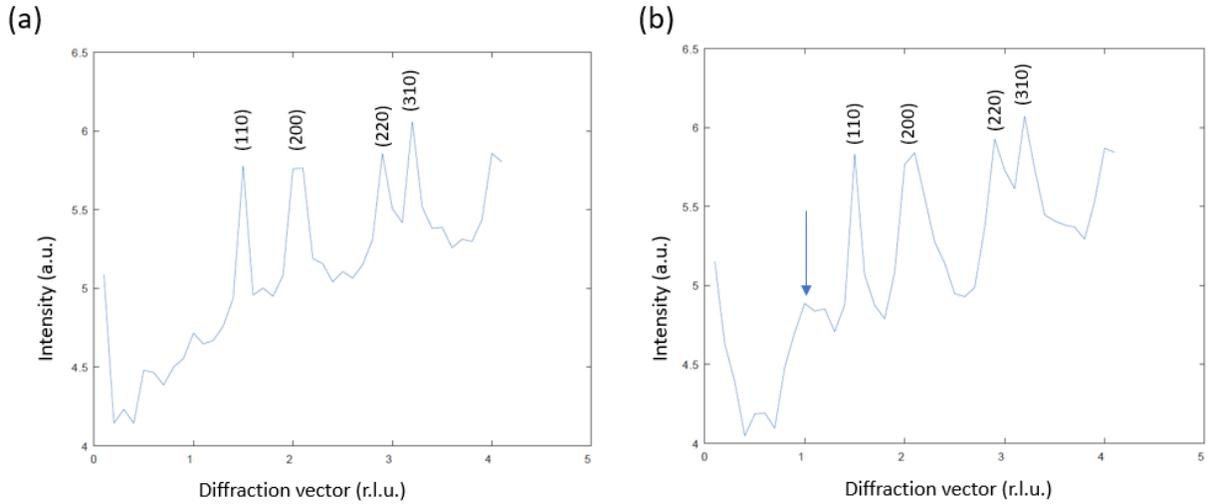

Fig. 6. Simulated powder diffraction pattern of MC/MD evolved R-HEA; (a) for system evolved at 800 K and (b) for system evolved at 100 K, respectively.

From the SRO parameter analysis in above sections it appears that there is tendency of segregation of Nb and W as next neighbors. Whereas there is strong tendency of the Mo-Ta pairs to order next to each other which resulted in a separate B2 ordering type scattering intensity in the simulated diffractions for the low temperature structures. In Fig. 7 some of the atomic structures of the MC/MD evolved systems for various temperatures are presented. The 10x10x10 supercell BCC structures are presented along with their periodic images for better visualization purpose. It can be observed that for the very low temperature evolved structures like 100 K and 300 K in Figs. 7a-b, there are separate patches of blue and green atoms, which are essentially Nb and W, respectively. This mutual segregation of Nb with W is supported by the large positive value of SRO parameter and positive $\Delta H$ of ordering discussed earlier. As the temperature rises to 700 K, there is increased randomness observed in the local distribution of atoms in Figs. 7c, d, f. In Fig. 7e, there is a cross section of the evolved structure for 100 K shown, which shows local patches of next neighbor ordering of the pink and white atoms. These are the local areas of Mo-Ta next neighbor B2-type ordering. The presence of these Mo-Ta ordering is also suggested from the large negative value of SRO parameter, large negative $\Delta H$ of ordering and the appearance of diffraction scattering peak in simulations. As the local chemical orderings appear mainly below 600 K, this HEA needs to be aged theoretically even at ambient temperature for long time to observe chemical ordering. But it is unlikely that at ambient temperature or below that there would be any thermal kinetic energy in the system for any diffusion to take place. These thermodynamic and kinetic factors put together gives the MoNbTaW HEA its exceptional phase stability for a very wide temperature range.

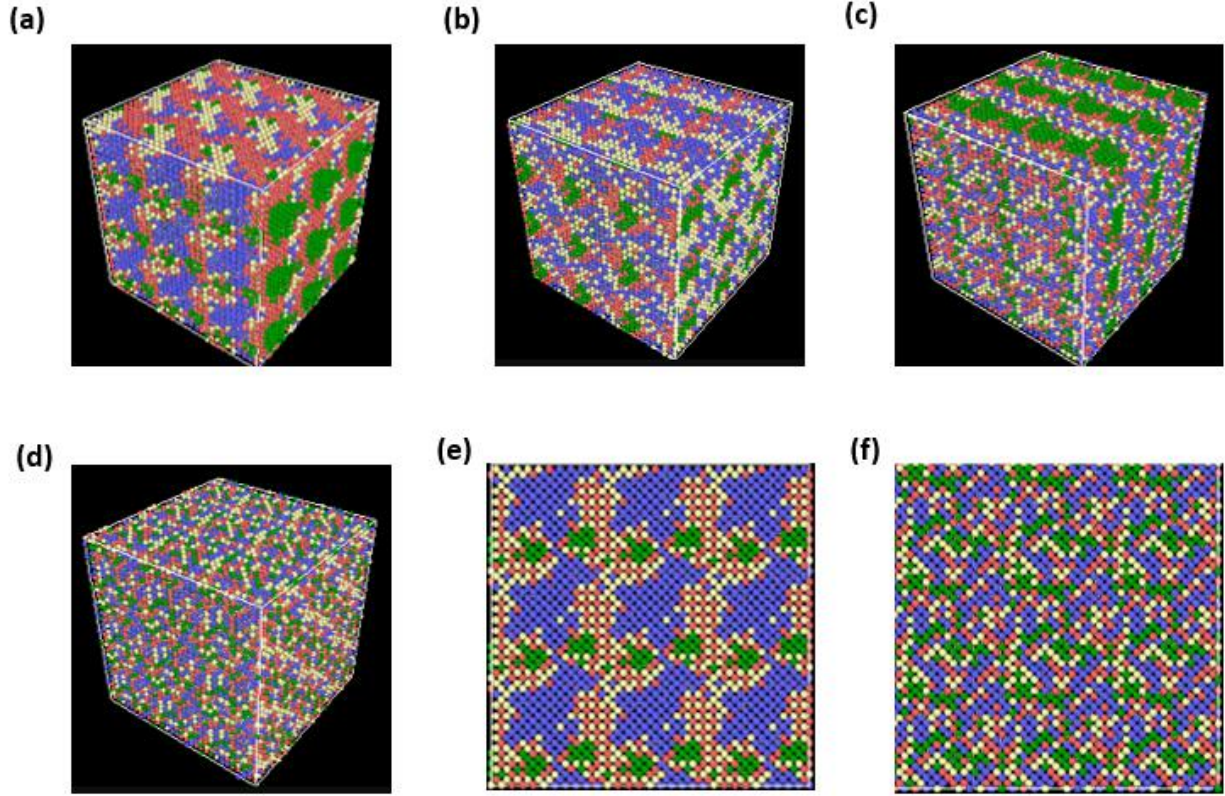

Fig. 7. Equilibrium nanostructures of the MC/MD evolved R-HEAs at various temperatures with perspective views; (a) evolved at 100 K, (b) at 300 K, (c) at 500 K, (d) at 700 K, respectively; (e) is cross-sectional nanostructure for 100 K and (f) for 700 K, respectively. Pink atoms are Mo, blue atoms are Nb, white atoms are Ta and green are W, respectively.

## 5. Conclusion

The conclusions of this work on the temperature dependent phase-stability of MoNbTaW R-HEA can be summarized into these following points.

1. Embedded atom method (EAM) potential based hybrid Monte Carlo molecular dynamics (MC/MD) can predict the important SRO/ SRC behavior of HEAs such as Mo-Ta and Nb-W pairs as also predicted by various other DFT based calculations.
2. The SRO transition temperature predicted by this EAM potential based study is similar to the DFT calculation based studies, which otherwise takes lot of computation resource.
3. The thermodynamic evolution of free energy suggests that the equilibrium phase evolutions are largely enthalpy driven for this R-HEA system.

4. The predicted SRO transition temperature of 700 K is in accordance with the experimental findings from synchrotron XRD and other DFT based calculations from the literature.
5. The evolved nanostructures at lower temperatures have presence of both short-range ordering (Mo-Ta) and short-range clustering (Nb, W) as expected from the Warren-Cowley parameters determined from the MC/MD simulations.
6. The developed methodology in this study can help accelerate new materials discovery by reducing the numbers of costly experiments to study the phases stability of multi-element alloys.

**Acknowledgement**

The authors would like to thank Mr. Ananth Krishnan, CTO of Tata Consultancy Services Ltd. for supporting this research. The authors are grateful for the access to the computing clusters of TCS Research, Pune. Any discussion with Prof. Walter Steurer of ETH Zurich, Switzerland is thankfully acknowledged.